\documentclass[aps,showpacs,preprintnumbers,floats,11pt]{revtex4}

\usepackage{graphicx}% Include figure files
\usepackage{dcolumn}% Align table columns on decimal point
\usepackage{bm}% bold math
\usepackage{epsfig}
\usepackage{amsmath}

\begin{document}

\title{On the momentum-dependence of $K^{-}$-nuclear potentials }
\author{
 L.\ Dang$^1$\footnote{E-mail: xiaoerqun@yahoo.com.cn },
 L.\ Li$^1$\footnote{E-mail: lilei@nankai.edu.cn}
 X.\ H.\ Zhong$^2$\footnote{E-mail: zhongxh@ihep.ac.cn}, and
 P. Z. Ning$^1$\footnote{E-mail: ningpz@nankai.edu.cn}
       }
\affiliation{
  $^1$Department of Physics, Nankai University, Tianjin 300071, China\\
 $^2$Institute of High Energy Physics,
       Chinese Academy of Sciences, Beijing 100039, China}
%\date{\today}

\begin{abstract}
 The momentum dependent $K^{-}$-nucleus optical potentials are obtained based on
the relativistic mean-field theory. By considering  the quarks
coordinates of $K^-$ meson, we introduced  a momentum-dependent
``form factor" to modify the coupling vertexes. The parameters in
the form factors are determined by fitting the experimental
$K^{-}$-nucleus scattering data. It is found that the real part of
the optical potentials decrease with increasing $K^-$ momenta,
however the imaginary potentials increase at first with increasing
momenta up to $P_k=450\sim 550$ MeV and then decrease. By comparing
the calculated $K^-$ mean free paths with those from $K^-n$/$K^-p$
scattering data, we suggested that the real potential depth is
$V_0\sim 80$ MeV, and the imaginary potential parameter is $W_0\sim
65$ MeV.
\end{abstract}

\keywords {momentum dependence optical potential,kaon-nucleus}

\pacs{21.65.+f, 21.30.Fe}

\maketitle

Recently kaon nuclear physics has been a hot topic of nuclear
physics, and kaon-nucleus interaction is a key point to many studies
on kaon. The information of $K^-$-nucleus interaction were obtained
from $K^{-}$- atomic and $K^-N$ scattering data. Since kaon is only
sensitive to the surface structure of nuclei in $K^{-}$- atomic,
predictions on $K^-$-nucleus interactions are very different for
different models (a very strong attractive real potential, $150\sim
200$ MeV, from the density dependent optical potential (DD) model
\cite{Frid,Frid1999}; and a much shallower one $\sim50$ MeV from the
chiral model \cite{Hirenzaki}).
 On the other hand, Sibirtsev \emph{et al.}
predicted the kaon-nucleus interaction has ``momentum dependence"
from $K^-N$ scattering data \cite{sib}. They have obtained a
momentum dependent potential in a dispersion approach at normal
nuclear density, the potential depth is about $140\pm 20 $MeV at
zero momentum, and decreases rapidly for higher momenta. In our
previous work \cite{zh}, we also found that the kaon nucleus optical
potential has strong momentum dependence by fitting the only
experimental data on the $K^-$-$C$, $K^-$-$Ca$ scattering at
$P_k=800$ MeV/c \cite{dmar}. We indicated that the depth of real
potential at the inner nuclei is $(45\pm 5)$ MeV at $P_k=800$ MeV/c,
which is much shallower than that at zero momentum in the RMF. We
shall here be concerned with discussing the momentum dependence of
$K^-$-nucleus interaction within the framework of RMF.

 In the usual RMF model, one cannot
obtain the correct momentum dependence of $K^-$-nucleus interaction,
and have to take the internal structure of kaon into account to
introduce a momentum-dependent ``form factor" \cite{dow}. When RMF
is extended to study KN interaction at the quark level, the same
approximation
 as made in the quark-meson
coupling (QMC) model\cite{kts, kts1} should be introduced: $\sigma$-
and $\omega$-mesons are exchanged only between the $u, d$ quarks or
their anti-quarks in $K$ meson, contributions from the $s$ quark are
ignored. In the following, we shall take kaon as a two quarks system
to introduce an exponential ``form factor",
$\exp(-P^2_k/4\kappa^2)$, which modify the couplings vertexes. In
fact, similar exponential ``form factor" has been widely adopted to
improve various calculations \cite{zhao,exf,exf1}.

\begin{figure}[ht]
\center
 \epsfig{file=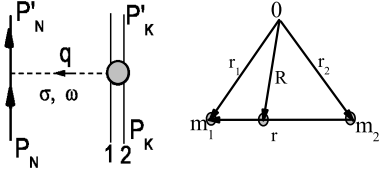,width=6.0cm} \caption{Feynman diagram for the KN interactions
 and internal coordinate of two quark system}
\label{opt}
\end{figure}

In the RMF, the $KN$ interactions are described  by exchanging
scalar meson $\sigma$ and vector meson $\omega$ \cite{zhong}. At the
quark level, $\sigma$- and $\omega$-mesons are exchanged between the
$u, d$ quarks or their anti-quarks. See Fig. \ref{opt}, the scalar
and vector couplings are

\begin{eqnarray}
\mathcal{L}_\sigma&=&g_{\sigma K}m_K \bar{K}   \sigma K,\\
\mathcal{L}_\omega&=&ig_{\omega K} \bar{K} \omega^{\mu}
\partial_{\mu} K+ H.C..
\end{eqnarray}
Replacing the scalar meson $\sigma$ and vector meson $\omega$ with
their plane wave form to get,
\begin{eqnarray}
\mathcal{L}_\sigma&=&g_{\sigma K}m_K\bar{K}  \left(a_\sigma e^{-i\textbf{q}\cdot \textbf{r}_2}\right)  K,\\
\mathcal{L}_\omega&=&ig_{\omega K}\bar{K}  \left(a_\omega
e^{-i\textbf{q}\cdot \textbf{r}_2}\varepsilon^\mu\right)
\partial_{\mu} K+ H.C.,
\end{eqnarray}
where $\textbf{r}_2$ represents the coordinates of $u/ d$ quarks,
with respect to the center of mass coordinates  $\textbf{R}$ and the
relative coordinates $\textbf{r}$ in the quark model. And we assume
$\textbf{q}=c\textbf{P}_k$ approximately as the transferred momentum
being proportional to $K^-$ incidence momentum. Thus one get
\begin{eqnarray}
e^{-i\textbf{q}\cdot \textbf{r}_2}=e^{-ic\textbf{P}_K \cdot
\textbf{R}}e^{\frac{\mu}{m_2}ic\textbf{P}_K \cdot\textbf{r}}.
\end{eqnarray}
On the harmonic-oscillator basis,  if the relative coordinates
 $\textbf{r}$ is rewritten in the second-quantized form, then
\begin{eqnarray}
e^{-i\textbf{q}\cdot \textbf{r}_2}=e^{-ic\textbf{P}_K \cdot
\textbf{R}}e^{\frac{c^2\textbf{P}^2_K}{4\alpha^2(m_u/\mu)^2}}e^{\frac{\mu}{m_2}ic\textbf{P}_K
\cdot\textbf{a}^{\dagger}}e^{\frac{\mu}{m_2}ic\textbf{P}_K
\cdot\textbf{a}}.
\end{eqnarray}

Finally, approximately we have
\begin{eqnarray}
\mathcal{L}_\sigma&\simeq &g_{\sigma K} m_K F(P^2_k)\bar{K} K
a_\sigma e^{-ic\textbf{P}_K \cdot \textbf{R}},\label{17}\\
\mathcal{L}_\omega&\simeq &ig_{\omega K}F(P^2_k)\bar{K}
\varepsilon^{\mu} a_\omega e^{-ic\textbf{P}_K\cdot \textbf{R}}
\partial_{\mu} K+ H.C.,\label{8}
\end{eqnarray}
with
\begin{eqnarray}
F(P^2_k)\equiv\exp[-P^2_k/(4\kappa^2)],
\end{eqnarray}
where $\kappa^2 \equiv \alpha^2(m_u/c\mu)^2, $ and
$P_k=|\textbf{P}_K|$.

 In the c.m. system of $K$-meson, replace $
a_\sigma e^{-ic\textbf{P}_K\cdot \textbf{R}}$ (
$a_\omega\varepsilon^{\mu}e^{-ic\textbf{P}_K\cdot \textbf{R}}$) with
$\sigma$ ($\omega^\mu$), Eqs.(\ref{17}, \ref{8}) can be rewritten
as,
\begin{eqnarray}
\mathcal{L}_\sigma&\simeq &g_{\sigma K}F(P^2_k)m_K \bar{K} K
\sigma\label{aa},\\
\mathcal{L}_\omega&\simeq &ig_{\omega K}F(P^2_k)
\left[\bar{K}\partial_{\mu}K-K\partial_{\mu}\bar{K}\right]\omega^{\mu}.\label{bb}
\end{eqnarray}

Comparing with the formulation without considering the quarks
coordinates of $K$-meson, it is obvious that an additional factor
$F(P^2_k)$ appears in the vertexes. Thus, the usual
 RMF Lagrangian for $KN$ interaction should be modified as
\begin{eqnarray}
{\mathcal{L}}_{\mathrm{K}}=&&
 \partial_{\mu}\bar{K}\partial^{\mu}K
 -m_K^2\bar{K}K-g_{\sigma K}m_{K}F(P^2_k)\bar{K}K\sigma
\nonumber\\
 &&-ig_{\omega K}F(P^2_k)\left[\bar{K}\partial_{\mu}K-K\partial_{\mu}\bar{K}\right]\omega^{\mu}\nonumber\\
&& +\left[g_{\omega K}F(P^2_k)\omega^{\mu}\right]^2\bar{K}K.
\end{eqnarray}
After a few simply deductions \cite{zhong}, we can obtain the real
part of the $K$-nucleus optical potential,
\begin{eqnarray}
\mathrm{Re}U
=&&\big[g_{\sigma\mathrm{K}}m_{\mathrm{K}}\sigma_{0}-2g_{\omega\mathrm{K}}E_{\mathrm{K}}\omega_0
-F(P^2_k)(g_{\omega\mathrm{K}}\omega_0)^2\big]\nonumber\\ && \cdot
F(P^2_k)/2m_K, \label{realp}
\end{eqnarray}
which refer to the $K$-meson three momenta by the form factor
$F(P^2_k)$. The $K^-$-meson energy $E_{\mathrm{K}}$ can also be
deduced from the RMF model \cite{zhong},
\begin{eqnarray}
E_\mathrm{K}=\sqrt{m_{\mathrm{K}}^2+g_{\sigma\mathrm{K}}F(P^2_k)m_{\mathrm{K}}\sigma_{0}+P_{k}^2}
-g_{\omega\mathrm{K}}F(P^2_k)\omega_0. \label{ek}
\end{eqnarray}
Here the coupling constants $g_{\sigma K}=$2.088 and $g_{\omega
K}=$3.02, which are used mostly in the RMF\cite{zhong}.

Up to now the anti-kaon absorption in the nuclear medium are
ignored, since imaginary potential cannot be obtained directly from
RMF. Similar to our previous work \cite{zhong}, we assumed a
specific form of the imaginary potentials:
\begin{eqnarray}
\mathrm{Im}
U=-f\cdot[F_2(P^2_k)]^2\cdot\left[\frac{E_{\mathrm{K}}}{m_k}W_{0}
\frac{\rho}{\rho_{0}}\right], \label{imgp}
\end{eqnarray}
where $ F_2(P^2_k)=e^{-P^2_k/(4\beta^2)}$, which is also introduced
to modify the imaginary potential (i.e. decay widths) as did in the
real part. For the decay width $\Gamma\propto \mathcal{M}^2\propto
g^2$, where $\mathcal{M}$ is the decay amplitude, and $g$ is the
coupling, the square of the ``form factor" $[F_2(P^2_k)]^2$ is
adopted. Besides, the phase space available for the decay products
should be considered \cite{d,Galaa}, which effects the imaginary
potentials (widths). Thus, a factor, $f$, multiplying imaginary
potentials Im$U$ is introduced in our calculations, as did in Ref.
\cite{zhong} (replace  Re$E$ with $E_\mathrm{K}$ of Eq. (\ref{ek})).
The factor $f$ can be assumed a mixture of 80\% mesonic decay and
20\% non-mesonic decay \cite{d,Galaa}, thus $f = 0.8f_1 + 0.2f_2$.
The imaginary potential parameter $W_0$, which is the depth of the
imaginary potential at zero momentum, is not determined well. By
fitting the $K^-$- atomic data, $W_0\sim 50$ MeV
\cite{Frid,Frid1999}, however, the predictions in Refs.
\cite{Shev,MOT} give a much deep value $W_0\sim 100$ MeV. In this
work, we shall discuss several cases for $W_0=50,65,80$ MeV.

\begin{figure}[ht]
\center
 \epsfig{file=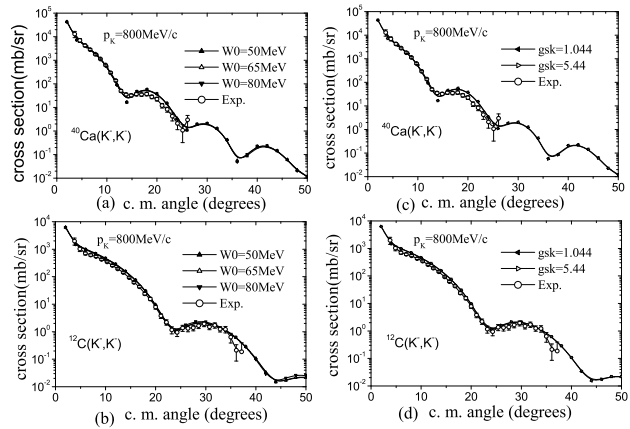,width=9.2cm} \caption{On the several cases of different real (c, d) and imaginary (a, b) potentials,
 the elastic differential cross section for $K^{-}$
scattering from $^{40}Ca$ and $^{12}C$ as functions of c. m. angles
at $p_{k}$=800 MeV/c are shown in Figs.(a, c) and (b,d),
respectively. The experimental data are from Ref. \cite{dmar}.}
\label{cross}
\end{figure}

\begin{figure}[ht]
\center
 \epsfig{file=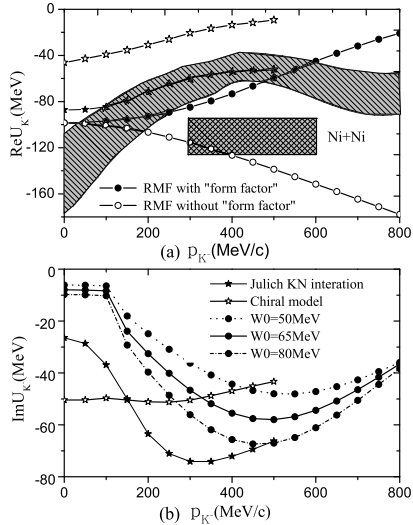,width=6.0cm} \caption{The real and imaginary anti-kaon optical potentials
  in normal nuclear matter based on our model ( solid dotted lines)
 and the other models are shown in Fig.(a) and (b), respectively.
The solid pentagons curves are the predictions of a meson-exchange
model with the J$\ddot{u}$lich $KN$ interaction \cite{a.ramos}. The
hollow pentagons curves are the results based on the lowest-order
meson-baryon chiral lagrangian (the anti-kaons and pions are dressed
self-consistently) \cite{laura}. The shadow region between two solid
curves is the results predicted by Sibirtsev and Cassing (SC model)
\cite{sib}, and the crossed rectangle indicate the results from the
analysis of $K^-$ production in $Ni+Ni$ collisions \cite{c5,wc}. }
\label{opt1}
\end{figure}

Finally, a momentum-dependent $K^{-}$-nuclear potential is obtained.
Naturally, we do not expect the naive quark model to give
appropriate values for the parameter $\kappa$ and $\beta$. In the
calculation, the parameters $\kappa$ and $\beta$ are determined by
fitting the $K$-nucleus scattering data. The experimental data of
the differential elastic cross sections for $K^{-}$-$^{12}C$ and
$K^{-}$-$^{40}Ca$ at $P_{\mathrm{K}}$= 800 MeV/c \cite{dmar} are
used to determine the parameters $\kappa=0.275$ GeV and $\beta=
(0.49, 0.44, 0.42)$ GeV (corresponding to $W_0=50,65,80$ MeV,
respectively). In Fig. \ref{cross}(a, b), according to the optical
potentials from equations (\ref{realp}) and (\ref{imgp}), the
experimental data are fitted very well. With the determined
parameters, we plotted the potentials as functions of kaon three
momentum $P_k$ at normal nuclear density in Fig. \ref{opt1}. The
real and imaginary parts are shown in Fig. \ref{opt1}(a) and (b),
respectively. The solid dotted lines are our calculations of Eqs.
(\ref{realp}, \ref{imgp}), the potentials predicted by the others
\cite{a.ramos,laura, sib} are also presented in the same figure.

From Fig. \ref{opt1}(a), we can see that our results on real
potentials decrease with increasing momenta, their varying
tendencies are in agreement with the other models
\cite{a.ramos,laura, sib}. The depths predicted by us are deeper
than those of chiral and J$\ddot{u}$lich models. Among these models,
chiral model gives much shallower real potential depths than the
other three models. The results of J$\ddot{u}$lich $KN$ interaction
and RMF model (with form factors) are compatible at $P_K< 600$ MeV,
which are almost in the possible region predicted by SC model. On
the other hand, the real potential based on the RMF without ``form
factor"
  is also shown in Fig. \ref{opt1}(a). It is obvious that $F_1(P^2_K)$
  has a great influence on the real potential, its corrections to the
  varying
   tendency of the real potentials are important. From Fig. \ref{opt1}(b), our results of the imaginary
potentials increase at first with increasing momenta up to
$P_k=450\sim 550$ MeV and then decrease, their varying tendencies
are in a similar way to the results of Ref. \cite{a.ramos}. There is
a flat for imaginary potential curve in the low energy $P_k< 100$
MeV region, which is referred to the factor $f_1=0$, indicates that
the total energy $(M_N+E_K)$ is less than the threshold of $\Sigma
\pi$, and the decay channel $NK\rightarrow \Sigma \pi$ is closed.

\begin{figure}[ht]
\center
 \epsfig{file=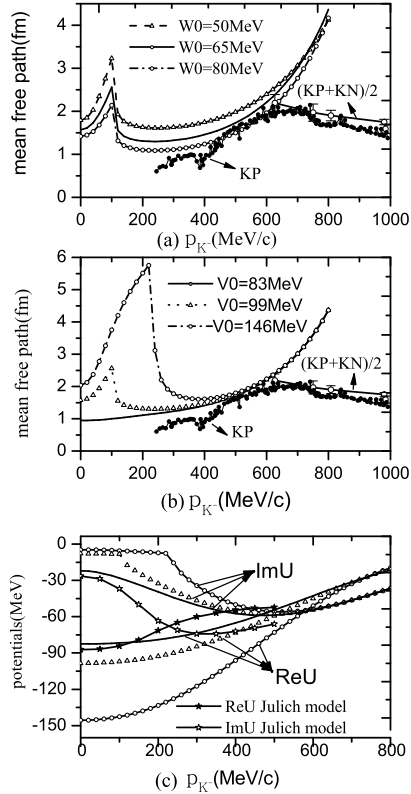,width=6.0cm} \caption{According to the different imaginary (real)
potentials, the $K^-$ mean free paths as functions of the incident
momenta in normal nuclear matter are shown in Fig.(a)(Fig.(b)). The
mean free path from the experimental $Kp$ and ($Kp+Kn$)/2 total
cross sections are also presented. Corresponding to the
$\lambda_\mathbf{K}$ in Fig.(b), the real and imaginary parts of
$K^-$
 optical potentials are also shown in Fig.(c).} \label{mfp0}
\end{figure}

However, $K^-$-nucleus elastic scattering data are not good enough
to test the validity of the physics contained in our model. Since
little experimental information came directly from the inner nuclei
for kaon, there are much uncertainties in both the real potentials
and the imaginary parts. The $K^{-}$ mean free paths (MFP) in
nuclear matter can be calculated with the determined
momentum-dependent potentials, which can also be estimated from the
experimental data of the total cross sections for $K^-p$ and $K^-n$
\cite{PDG}. By comparing the results in two different approaches, we
expect to find more constraints on the $KN$ interactions.

The details of how to calculate a particle's MFP are given in our
previous work \cite{wang}, only the formula of
$\lambda_{\mathrm{K}}$ is given here,
\begin{eqnarray}
\lambda_{\mathrm{K}}=\frac{1}{2\sqrt{m_\mathrm{K}\cdot[B^{2}+(\mathrm{Im}
U)^{2}]^{\frac{1}{2}}
 -m_\mathrm{K}\cdot B}}, \label{lambda1}
\end{eqnarray}
where, $B\equiv E_\mathrm{K}-m_\mathrm{K}-\mathrm{Re}
U+(E_\mathrm{K}- m_\mathrm{K})^{2}/2m_\mathrm{K}$. On the other
hand, the MFP of $K^-$ is related to the $K^-p$/$K^-n$ scattering
data by a simple relation $\lambda=1/\rho\overline{\sigma}$, where
$\bar{\sigma}=(\sigma_{Kn}+\sigma_{Kp})/2$ is the average of total
$K^-n$ and $K^-p$ cross sections. There are some $K^-p$ scattering
data in the range of $240<P_k<1000$ MeV, and a few data for $K^-n$
scattering in $600<P_k<1000$ MeV. Thus only the MFP data from
$\bar{\sigma}=(\sigma_{Kn}+\sigma_{Kp})/2$ in the latter region can
be compared. The results of two approaches are shown in Fig.
\ref{mfp0}.

 From the figure, we find that
 $\lambda_p$ ($=1/\rho\sigma_{Kp}$) is a little larger
than $\lambda$ ($=1/\rho\overline{\sigma}$), and we assume
 $\lambda_p\simeq \lambda$ in the region of $P_k<600$
MeV. There is a ``peak" in each of our calculated curves of the MFP,
which is referred to the factor $f_1=0$, corresponds to the position
of $M_N+E_K=M_\Sigma +M_\pi$. By comparing the results of different
imaginary potential parameter $W_0$ with the $\lambda_p$, and
considering $\lambda
> \lambda_p$ and $\lambda \simeq \lambda_p$, we think the most possible  imaginary depth
parameter $W_0$ should be $\sim 65$ MeV.

If one takes the coupling constant $g_{\sigma K}$ to be a free
parameter, the different real potential depths can be obtained by
adjusting $g_{\sigma K}$. With $W_0=65$ MeV (the corresponding form
factor parameter $\beta$ = 0.44 GeV), the mean free paths are
calculated for the different real potential depths $V_0=83,\ 99,\
146$ MeV (the corresponding coupling constant $g_{\sigma K}$=1.044,
2.088, 5.44), which are shown in Fig. \ref{mfp0}(b). The
corresponding real and imaginary parts of the optical potentials are
also shown in Fig. \ref{mfp0}(c). The form factor parameters of the
real potential ( $\kappa$ = 0.285, 0.255 GeV correspond to
$g_{\sigma K}$=1.044, 5.44, respectively) are determined by fitting
the $K^-$ nucleus scattering data (which are shown in Fig.
\ref{cross}(c, d)). From Fig. \ref{mfp0}(b), we can see that our
calculations of $V_0=83$ MeV are most compatible with the
$\lambda_p$ from $K^-p$ scattering data. And in Fig. \ref{mfp0}(c),
with $V_0= 83$ MeV and $W_0 = 65$ MeV, both the real and imaginary
potentials (the solid curves) of our calculations are very close to
those of J$\ddot{u}$lich $KN$ interactions \cite{a.ramos} (the star
curves) in the range of $P_k<100$ MeV. It is interesting that the
recent experiment also indicated that the in-medium $K^-N$ potential
depth is about $\sim80$ MeV at normal nuclear density \cite{expk}.

 As a whole, with the constraints of the MFP from KN
scattering data, we predicted that the real potential depth is
$V_0\sim 80$ MeV, and the imaginary parameter $W_0\sim 65$ MeV. One
point must be emphasized, if the above results about the potential
depths are right, according to our calculations on $K^-$-nuclei in
\cite{zhong}, the sum of the half widths of the 1s and 1p states are
larger than their separations in $K^-$-nuclei. In other words, no
discrete $K^-$ bound states in the $A\geq 12$ nuclei can be found in
experiments.

 In conclusion, the
momentum dependence of $K^-$ nucleus potentials have been studied in
the framework of RMF theory. We think that the interior structure of
a kaon may be one of the origin of the momentum dependence, and
introduce a ``form factor" to correct both the real and imaginary
parts of the potential. It is found that  the real part of the
optical potentials decrease with increasing $K^-$ momenta, however
the imaginary potentials increase at first with increasing momenta
up to $P_k=450\sim 550$ MeV and then decrease. The effects of the
exponential form factor on real and imaginary potentials are
important. Analyzing several cases on both the real and imaginary
potential depths, we predicted that the real potential depth is
$V_0\sim 80$ MeV, and the imaginary parameter $W_0\sim 65$ MeV with
the constraints of the MFP from KN scattering data.

X.H. Zhong would like to thank Prof. H.Oeschler for useful
discussions. This work was supported, in part,  by the Natural
Science Foundation of China (Grand No. 10575054), China Postdoctoral
Science Foundation, and the Institute of High Energy Physics, CAS.

\end{document}